\documentclass[a4paper,11pt]{article}
\usepackage{jheppub}

\usepackage{amsmath}
\usepackage{graphicx,tabularx}
\usepackage{url}
\usepackage{footmisc}
\usepackage{amsfonts}
\usepackage{cancel}
\usepackage{color}
\usepackage{multirow}
\usepackage{slashed}
\usepackage{comment}
\usepackage{appendix}
\usepackage{soul}
\newcommand{\fslash}{\hspace{-0.26cm} /}

\definecolor{Red}{rgb}{1,0,0}

\title{$D^*$ meson production in jet from combination of charm quark with light one}

\author[a]{Chuanhui Jiang,}
\author[b,1]{Honglei Li,\note{Corresponding author.}}
\author[a]{Shi-Yuan Li,}
\author[c]{Shufen Liu,}
\author[b]{Xinyue Yin}

\affiliation[a]{ Department of Physics, Shandong University, Jinan Shandong 250100, China}
\affiliation[b]{School of Physics and Technology, University of Jinan, Jinan Shandong 250022, China}
\affiliation[c]{Tai'an Eco-environmental Monitoring Station of Shandong Province, Tai'an Shandong 271000, China}
\emailAdd{jiangch@mail.sdu.edu.cn,sps\_lihl@ujn.edu.cn, lishy@sdu.edu.cn, liushf\_tae@163.com, yinxinyue@mail.ujn.edu.cn}

\abstract{
In the framework of the perturbative Quantum Chromodynamics factorization, the cross section of the heavy meson production via the combination of a heavy quark with a light one can be factorized to be  the convolution of  the combination matrix element, the light quark distribution function, and the hard partonic sub-cross section  of the heavy quark production. The partonic distribution and the combination matrix element are  functions of a scaling variable, respectively, which is the momentum fraction of the corresponding quark with respect to the heavy meson.  We studied the $D^{*\pm}$  production in jet via combination in pp collision at the LHC. Our calculation can be summed with the fragmentation contribution, and the total result is comparable with the experimental data. The combination matrix elements can be further studied in  various hadron  production processes.
  }
\begin{document}
\maketitle

\section{Introduction}\label{sec:intro}

Heavy flavor production in high energy collision provides the crucial test of Quantum Chromodynamics. While the accuracy of the test relies on the nonperturbative process of hadronization.  The partons in the early stage after the hard collisions can be measured as jets reconstructed by the final hadrons. It is an interesting measurement to study the hadronization through the jet structure, especially the hadron distributions inside jets. The ATLAS collaboration have reported the $D^{*\pm}$ meson production in jets from pp collision recently~\cite{ATLAS:2011chi}. The Monte Carlo predictions in the traditional framework of the fragmentation fail to describe the data for the $D^{*\pm}$ mesons carrying a small fraction of the jet momentum. Studies for the modified fragmentation functions including the high order corrections have been proposed to eliminate the discrepancy~\cite{Chien:2015ctp, Anderle:2017cgl}. They suggested to enhance the gluon fragmentation function  by a factor of 2, then the calculation leads to a better agreement with the data. On the other hand, it has long been recognized that hadrons with modest momenta where the multiplicity is large get contributions from (re)combination mechanism. In this paper we investigate the contribution from  the combination of charm quark with the light one nearby in phase space to produce the $D^*$ meson. The  underlying  events can provide the sources for the light quark to be combined.

Quark combination mechanism and practical models  have been widely studied for a very long time in various scattering processes. Among all the studies, the most relevant ones for this work are those from the viewpoint of perturbative Quantum Chromodynamics(pQCD)~\cite{Mueller:1985wy,Amati:1979fg}. Some years ago, we proposed the pQCD  factorized formulation for a heavy quark  produced from the hard scattering combining with a light one from the background environment to produce the heavy meson~\cite{Li:2005hh}. The obtained  cross section   of  the inclusive scattering  process $A+B \to M_Q +X$ (where  $M_Q$ denote the  the produced heavy  meson)  is the convolution of the hard sub-cross section of the heavy quark production $A+B \to Q \bar{Q}+X$,  the combination matrix element and the parameter (matrix element) corresponding to the  parton distributions in the background. The combination matrix element and the parton distribution of the background are expectation values of field operators on certain particle states, so they are model-independent and process-independent.  The combination matrix element describes the probability of a heavy quark and a light one of specified momenta to form a heavy hadron. Within the same factorization scheme, it can appear in other  more `simple and clean' processes, such as $e^+ e^-$ annihilation, etc.,  so it can be extracted from the experiments. The process-relevant factor is the spectrum of the heavy quark, which  can be calculated by pQCD. Now full NLO corrections are available~\cite{Cacciari:2003uh, Frixione:2010ra}. We suspected that this mechanism could be globally studied in various scattering  processes,  to quantify the universal  combination matrix elements. Such a study et vice verse could play the r\^{o}le  as a scaling probe on the background parton distribution, e.g., that of the quark gluon plasma in the relativistic  heavy ion collisions especially after the universal combination matrix elements are fixed~\cite{Li:2005hh}. Now the available data for charm meson distributions in a jet from pp scattering~\cite{ATLAS:2011chi} provide the good opportunity to study the combination mechanism as well as to probe the light quark distributions from the underlying events.

The outline of this paper is: In Section~\ref{sec:formucom}  we introduce the factorization framework and derive  the  cross section formula to describe the above combination mechanism. In Section~\ref{sec:num} we calculate the $D^{*\pm}$ distribution in jet $R(p_{T},z)$. The combination matrix element and the light quark distribution are yet beyond available from experiment now. We study the sensitivity of the $R(p_{T},z)$ to them. With some physical conjectures, we calculate the $D^{*\pm}$  production in jet via combination in pp collisions at the LHC. It is sumed with the fragmentation contribution, and the total result is comparable with the experimental data~\cite{ATLAS:2011chi}.  In Section~\ref{sec:disc}  we discuss some properties of the  combination mechanism studied in this paper and its possible  application to $J/\psi$-pair production.

%%%%%%%%%%%%%%%%%%%%%%%%%%%%%
\section{Cross section for combination of heavy quark  with light one}\label{sec:formucom}

\begin{figure}[thb]
%\label{cvf}
	\centering
	\begin{tabular}{cccccc}
		\scalebox{0.60}[0.6]{\includegraphics{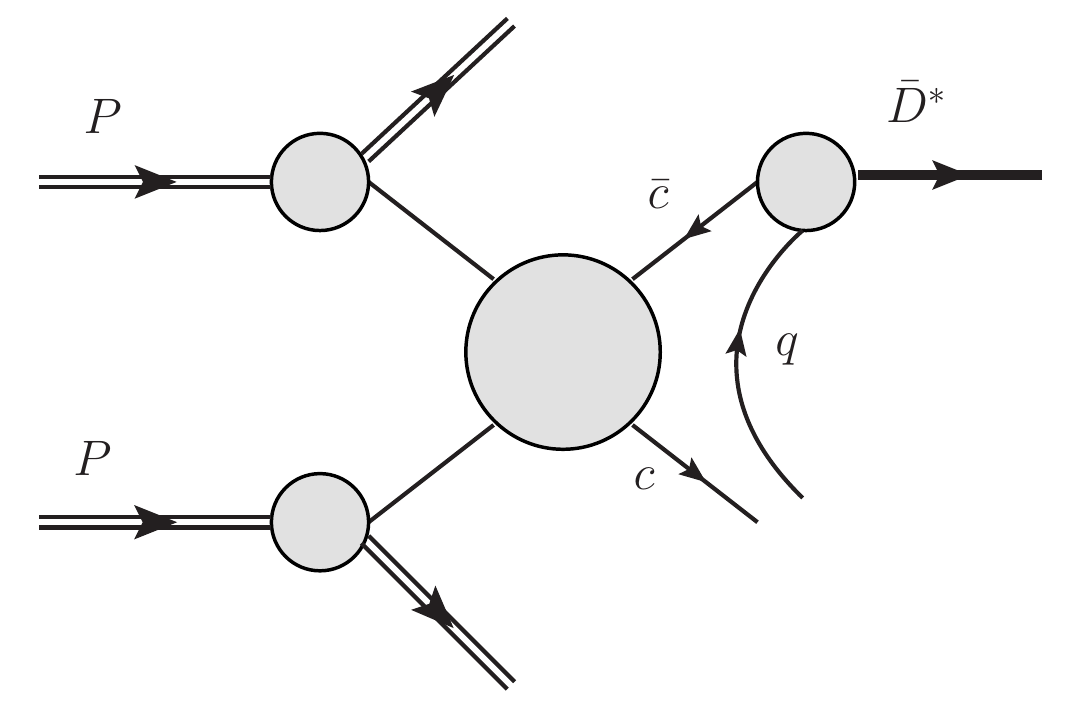}}		
	\end{tabular}
	\caption{$\bar{D}^{*}$ production at the proton-proton collider.}
	\label{fig:0}
\end{figure}

 In  this section, we take the hadronic collision $A+B \to \bar D^* + X$ as an  example, see Fig.~\ref{fig:0}. Here $\bar D^*$ refers to  any  anti-charm meson. The formula is the same when apply to the charm quark case. One studies prompt $D^*$ rather than prompt $D$ is because the former does not need to consider the decay contribution. We only consider the contribution of $\bar c+q$ to $\bar D^*$, other possible combination processes  such as $\bar c+g$ to $\bar D^*$ are assumed negligible. The $X$ includes the associated produced $c$ quark and all the other particles from the incident particles $A$, $B$  interaction.  We  obtain the  invariant inclusive differential production cross section $2E \frac{d {\cal \sigma }_C^{AB}}{d^3 \bf K}$ of $\bar D^*$. The  subscribe $C$ of  $\cal \sigma $  denotes  the combination process. $(E, \bf {K})$ is the 4-momentum of  $\bar D^*$.  The production of charm mesons can be treated in the same way as the anti-charm mesons. To describe the light quark to be combined, we should find ways to represent the `external particle source'.  We introduce  an external vector $V^{\mu}_{ext}$ field which, together with quark field operators, appears in the matrix element corresponding  to  the quark distribution. The external field is  proportional to $n^{\mu}=(0, 1, {\bf 0}_{\perp})$. This can be understood as gauge fixing and is easy to factorize the Dirac indices.

From the above discussions, the interaction Hamiltonian for quark and gluon fields is extended as~\cite{Qiu:2003pm}:
\begin{equation}
\label{intham}
H_I= \bar \Psi ({  A \hspace{-0.26cm}  \slash}+ { V \fslash}) \Psi.
\end{equation}
 Here $A$ is the normal gluon field and $V$  is the external field. The strong  coupling constant $g_s$ is absorbed into the gauge fields. In this paper, wherever we write the gauge field obviously, we always adopt  this convention. If  we assume that the distribution functions of the  intrinsic heavy flavours in the initial  state $AB$ are vanishing, the lowest order contribution for the combination process comes from $O(g_s^3)$ in the perturbative  expansion of S-matrix:
\begin{eqnarray}
\label{smatrics}
 S^{(3)}& = &\frac{(-i)^3}{3 !}C^2_3 \int d^4 x_1 d^4 x_2 d^4 x_3 {\bf T} \bar \Psi(x_1)  A\fslash (x_1) \Psi(x_1)  \nonumber \\
&\times &\bar \Psi(x_2)  A\fslash (x_2) \Psi(x_2) \bar \Psi(x_3)  V\fslash (x_3) \Psi(x_3),
\end{eqnarray}
where summation on colour and flavour indices as well as the indices in spinor space are indicated.

Now we  take the annihilation partonic process $q \bar q \to c \bar c$ as an example to illustrate the derivation, while the total result can be obtained by summing all kinds of partonic processes.  Let the corresponding terms of the  Wick expansion act on the initial  state $|A B>$  and final state $| \bar D^* X>$, employing the space-time translation invariance, we can isolate the $\delta$ function corresponding to  the total energy-momentum conservation and get the T-matrix element. The  cross section  is
\begin{eqnarray}
\label{jiemian1}
\sigma&= & \frac{(4 \pi \alpha_s)^2}{4 F}  \sum \limits_{\bar D^* X}  \int d^4x_1 d^4x_2 d^4x_3 d^4x_4 d^4x_5  \int{ \frac{d^4 q}{(2 \pi)^4} \frac{d^4 q'}{(2\pi)^4}} \frac{1}{q^2 q'^2} e^{-iq(x_1-x_2)} e^{iq'(x_4-x_5)} \nonumber \\
 & \times &
< A B| \bar \psi(x_4) T^{c_4} \gamma^{\mu_4} \psi(x_4)  \bar \psi(x_3) \not V(x_3) \psi(x_3) \bar \Psi(x_5) T^{c_4} \gamma_{\mu_4}  \Psi(x_5)|\bar D^* X> \nonumber \\
 &\times &
< \bar D^* X|  \bar \Psi(x_2) T^{c_1} \gamma_{\mu_1} \Psi(x_2)  \bar \psi(0) \not V(0) \psi(0) \bar \psi(x_1) T^{c_1} \gamma^{\mu_1}  \psi(x_1)|  A B>.
\end{eqnarray}
In the above equation, $T^c$ is one half of the Gell-Mann Matrix, $4 F$ represents the incident  flux factor and discrete quantum number average.  Summation on repeated indices is indicated. The capital  $\Psi$ is for the heavy quark field and the lower case $\psi$ for  light quark fields.

In the following the cross section is simplified  in the framework of collinear factorization:
\begin{equation}
\label{jiemian2}
\sigma=\frac{(4\pi \alpha_s)^2 C}{4 F} \int \frac{d^4 q}{(2\pi)^4} W^{\mu \nu} \frac{1}{q^4} D_{\mu \nu},
\end{equation}
with
 \begin{eqnarray}
 \label{wmunu}
 W^{\mu \nu}& = & \int d^4 x_4 e^{-iqx_4} < A | \bar \psi^{\alpha_{4}} (x_4) \psi^{\beta_{1}}(0)| A> <B| \psi^{\beta_4}(x_4) \bar \psi^{\alpha_1}(0)|B> \nonumber \\
&\times &  \gamma ^{\mu}_{\alpha_4 \beta_4} \gamma^{\nu}_{\alpha_1 \beta_1} + ( A  \leftrightarrow B).
\end{eqnarray}
This is the same $W^{\mu \nu}$  for Drell-Yan process, which  gives the distribution of initial partons. We have employed  the translation invariance and integrated on $x_1$, which gives $q=q'$.

On the other hand,
\begin{eqnarray}
\label{dmunu}
D_{\mu \nu}&=&\int \frac{d^3K}{(2 \pi)^3 2E} \frac{d^3k'}{(2\pi)^3 2E'} \int d^4 x_2 d^4 x_3 d^4 x_5 e^{-ik_{\bar c}x_2} e^{ik_{\bar c}x_5}\nonumber\\
&\times &(\gamma_\mu (\not k'+ m) \gamma_\nu)_{\alpha_5 \beta_2} <0|(\bar \psi(x_3) \not V(x_3))^{j_3}_{\beta_3}|X_h>\sum \limits_{ X_h}<X_h|(\not V(0) \psi(0))^{j_0}_{\alpha_0} |0> \nonumber\\
& \times &<0|\psi^{j_3}_{\beta_3}(x_3) \bar \Psi^{j}_{\alpha_5}(x_5)|\bar D^*><\bar D^*|\Psi^{j}_{\beta_2}(x_2) \bar \psi^{j_0}_{\alpha_0}(0)|0>.
\end{eqnarray}
To get the above expression, we have written the total final state produced  in the $A B$ collision as $|\bar D^* X>= |\bar D^* \tilde{X} c X_h>$, where the $\tilde X$ represents all the particles except the $c\bar c$ produced by the hard interaction and $X_h$ for the assemble of particles  produced by all the interactions  except the above hard one.  We have the corresponding field acting  on the $c$ quark final  state with momentum $k'$,  $k_{\bar c}=q-k'$. The summation on $\tilde{X}$  has been  eliminated  by the completeness condition. In Equation~(\ref{jiemian2}), $C$ is the colour factor of the partonic diagram except the external leg combined into $\bar D^*$. The colour part will be clarified following.

The $W^{\mu \nu}$ can be conventionally written as
\begin{eqnarray}
\label{wmnfac}
\int \frac{d^4 q}{(2 \pi)^4} W^{\mu \nu}&= & \int dr_1 dr_2 \int \frac{d \lambda_1}{2 \pi} \frac{d \lambda_2}{2 \pi} e^{-ir_1 \lambda_1} e^{-i r_2 \lambda_2} <A| \bar \psi(y)\frac{\gamma^+}{2P^+_1}\psi(0)|A> \nonumber\\
&\times &  <B| tr(\frac{\gamma^-}{2P^-_2}\psi(y) \bar \psi(0))|B> tr(\frac{\not P_1}{2} \gamma^\mu \frac{\not P_2}{2} \gamma^\nu),
\end{eqnarray}
which is the intended factorized form. Some of the above  variables are: $\lambda_1=P_1^+y_-, \lambda_2 =P_2^-y_+ $, $y=(y^+, y^-, {\bf 0}_\perp)$, $ Q^2 \equiv q^2 =r_1 r_2 s$, $s$ is the center of mass frame (c.m.s) energy  for the $AB$ system whose partons collide and produce the heavy quark pair.

The factorization for $D_{\mu \nu}$ is more complicated. It  can be written as:
\begin{eqnarray}
\label{dfac3}
& & \int \frac{d^3 K}{(2 \pi)^3 2E} d \xi K^+ d \xi_l K^+ (2 \pi) \delta(k'^2-m^2) (2 \pi)^4 \delta^4(k_l + k_{\bar c} -K)\nonumber \\
&\times &tr(\gamma_\mu (\not k'+m) \gamma_\nu \frac{\not{K}}{2})|_{k'=q-\xi K^+} \nonumber\\
&\times &\int d^4 x_3 e^{-i k_l x_3}|_{k^+_l = \xi_l K^+} <0| (V(x_3) \bar \psi(x_3))_{j_0}|X_h>\sum \limits_{X_h}<X_h| \frac{\gamma^+}{2 K^+}(\psi(0) V(0))_{j_0}|0>\nonumber \\
& \times &\frac{1}{9} \int \frac{d^4 k_{\bar c}}{(2\pi)^4} \frac{d^4 k_{l}}{(2\pi)^4} \delta(\xi-\frac{k^+_{\bar c}}{K^+}) \delta(\xi_l-\frac{k^+_{l}}{K^+}) \int d^4 x_2 d^4 x_5 e^{-i k_{\bar c} x_2} e^{i k_{l} x_5} \nonumber \\
& \times & <0| tr(\frac{\gamma^+}{2 K^+} \psi^{j_0}(x_5) \bar \Psi^{j}(0))|\bar D^*>
<\bar D^*| tr ( \frac{\gamma^+}{2 K^+} \Psi^{j}(x_2) \bar \psi^{j_0}(0))|0>.
\end{eqnarray}
In the above equation, the colour indices in the partonic final states and the distribution functions  are summed. So the colour indices in the combination matrix elements should be averaged ($\frac{1}{9}$), which is similar  as the case in fragmentation function. We do not separate the colour-singlet or the colour-octet contribution in the matrix elements, but sum them together. The reason is that  the partonic cross section is the same for the colour indices $j$  belonging  to $\underline{1}$ or $\underline{8}$ states (since the other parton comes from an un-correlated source), and that for the parton distribution, it should be the same whether $j_0$ belong to  $\underline{1}$ or $\underline{8}$ states. We have taken the external field proportional to $n^\mu$, which select only the ``+'' component since $\not{n}  K \fslash ^- \not n = \not n  {\bf K}\fslash  _\perp \not n=0$. In the part corresponding to the partonic sub-process, for the sake of factorization, we have done the collinear expansion for the momentum of the $\bar c$  along the ``+''  component of the momentum of $ \bar D^*$.  Here one notices  that the coordinate system is different from that for initial states $W^{\mu \nu}$. The z-axis direction is along  the momentum of the anti-charm meson.

In Equation (\ref{dfac3}), we write the combination matrix element (Row 4, 5) formally to be analogous to that in other processes (see following discussions).  $\xi$ and  $\xi_l$ seem not restricted to be $\xi+\xi_l=1$. However,   the $\delta$ functions in the first row sets  the restriction. At the same time, the integral in the combination matrix elements $\int{d^4 k_l}$ acts on the matrix element corresponding to the quark distribution  represented by the external field (Row 3)  as well as the $\delta$ function in the first row. These show that we have not finished the factorization. To get the factorized form, we notice
\begin{eqnarray}
&~ & \int \frac{d^3 K}{2E}\delta^4(k_l + k_{\bar c} -K) \nonumber \\
%&=& \int  d^4 K \delta^4 \delta(K^2-M^2)\\
%&=& \int d K^+ d K^- d^2 K_\perp  \delta^+ \delta^- \delta^2_\perp \delta(2 K^+K^- -K^2_\perp -M^2)\\
% integarte the last \delta with K^+
&= & \int \frac{dK^- d^2K_{\perp}}{2K^-}\delta^+( k^+_l + k^+_{\bar c} -K^+)\delta^- \delta^2 _\perp \nonumber \\
% because \int dK^- d^2K_{\perp}  \delta^- \delta^2_\perp=1, we take its place b%y anothe 1
%\int \frac{d^3 K}{2E} 2E \delta^3 ({\bf k}_{\bar c}+{\bf k}_l -{\bf K})=1
&= & \int \frac{d^3 K}{2E} 2E \delta^3 ({\bf k}_{\bar c}+{\bf k}_l -{\bf K}) \frac{1}{2K^-}\delta^+( k^+_l + k^+_{\bar c} -K^+).
\end{eqnarray}
Let the 3-dimension $\delta$ function  absorbed into the combination matrix element, we get the `restricted' matrix element or the dimensionless combination function:
\begin{eqnarray}
 \tilde{F}(\xi, \xi_l)&=&\frac{1}{9} \int \frac{d^4 k_{\bar c}}{(2\pi)^4} \delta(\xi-\frac{k^+_{\bar c}}{K^+})\frac{d^4 k_l}{(2\pi)^4} \delta(\xi_l-\frac{k^+_{l}}{K^+})  \int d^4 x_2 d^4 x_5 e^{-i k_{\bar c} x_2} e^{i k_l x_5} \nonumber \\
&\times & <0| tr(\frac{\gamma^+}{2 K^+} \psi^{j_0}(x_5) \bar \Psi^{j}(0))|\bar D^*>
<\bar D^*| tr ( \frac{\gamma^+}{2 K^+} \Psi^{j}(x_2) \bar \psi^{j_0}(0))|0> \nonumber \\
&\times & 2 E \delta^3 ({\bf k}_{\bar c}+{\bf k}_l -{\bf K}).
\label{eq:ft}\end{eqnarray}

The integral of $\xi_l$ in the first row of Equation (\ref{dfac3}) have given $\xi+\xi_l=1$. $\int d\xi  \delta(k'^2-m^2)$  gives important result:  $\xi=\frac{Q^2}{2K\cdot q}$, \footnote{This relation is exact for $Q$ and $K$ to be infinite while $z$ fixed. It is a good approximation when  the anti-quark to be combined into the heavy meson  is on mass shell in the partonic processes. In this case,  the four momentum of the anti-quark is $(\xi K^+, \frac{m^2}{2\xi K^+}, {\bf 0}_\perp)$. In the heavy meson rest frame,  it is easy to get $\xi =\frac{\sqrt{m^2+(mv)^2}+mv}{M} \simeq \frac{m}{M} $. Go back to the initial parton c.m.s, using this approximation for the on-shell condition  $(q-k)^2=m^2$, we can get the relation. So, it is not just an approximation by taking $M=m_c=0$ in $K \cdot q$.}  which is analogous to the Bjorken scaling variable. This means that for certain partonic c.m.s energy $Q$, the heavy meson with momentum $K$ just comes from the heavy quark with momentum fraction $\xi$ combined with light quark with momentum  $k_l=(1-\xi)K$, hence only {\it probe} this light quark by the combination mechanism.  Such a conclusion does not depend on the special forms of the derivation in this paper. In fact, the relation $\xi=\frac{Q^2}{2K\cdot q}$ is set by the on shell condition of the heavy quark   associatively produced  with the one to be   combined into the final state heavy hadron. Just like the DIS process, this is a physical condition which should be respected by any  special forms  of derivation.

The  cross section  now can be written as
\begin{eqnarray}
\label{cs11m}
%\begin{array}{ll}
2 E \frac{d \sigma_C}{d^3 K}&=& \frac{1}{4 F} \sum \limits_{ab} \int{dr_1 dr_2} 2 f^a_A(r_1) 2 f^b_B(r_2) \nonumber \\
& \times & | \tilde{\cal M}_{ab} | ^2  \frac{1}{\xi} \frac{(2 \pi)^2}{(2M)^2}  2 P (\xi_l) \tilde{F}(\xi, \xi_l) | _{\xi+\xi_l=1}.
%\end{array}
\end{eqnarray}
Here $f^a_A (r_1)$ and $f^b_B(r_2)$ are parton distributions with momentum factions $r_1$, $r_2$. $M$ is the mass of the heavy meson. $| \tilde{\cal M}| ^2$ refers to the invariant amplitude square including all the coupling constant and colour factors for the partonic process $ab \to \bar c +x$ (where the momenta of external legs are modified and $x$ to be considered as one particle). For example,  for $q \bar q \to \bar c +x$, to the lowest order,  $| \tilde{\cal M}| ^2$  is
\begin{equation}
\label{partoniccs}
(4 \pi \alpha_s)^2 C' tr(\frac{ P \fslash _1}{2} \gamma^\mu \frac{ P\fslash _2}{2} \gamma^\nu) \frac{1}{q^4} tr(\gamma_\mu (\not k'+m) \gamma_\nu \frac{ K \fslash }{2}).
\end{equation}
$C'$ is the colour factor. Though the quark mass term is vanishing, we keep it to show the origin of the formula. $P(\xi_l)$ can be understood as the distribution function of the parton (probed by the heavy quark).  $P(\xi_l)$ is also dimensionless:
\begin{equation}
\label{litd}
\frac{1}{2} \int{d^4 x_3} e^{-ik_lx_3} | _ {k_l^+=\xi_l K^+} \sum \limits_{X_h} <X_h| tr (\psi(0) V(0)|0> <0|V(x_3)\bar \psi(x_3) \frac{\gamma^+}{2 K^+})|X_h>.
\end{equation}
This can be understood as the expectation value on the  state representing an assemble of  particles produced in the $A B$ collision denoted by $|X_h>$, e.g., those from the underlying events.

We get the cross section of the  production  of $\bar{D}^*$  in the combination process:
\begin{equation}
\label{cossnum}
%\begin{array}{ll}
2 E \frac{d {\sigma_C}}{d^3 K}= \sum \limits_{ab} \int{dr_1 dr_2}  f^a_1(r_1)  f^b_2(r_2)
\frac{d \hat{\sigma}_{ab}}{d {\cal I}} \frac{1}{\xi^2} \frac{(2\pi)^2}{(2M)^2}  P(\xi_l) \tilde{F}(\xi, \xi_l)| _{\xi+\xi_l=1}.
%\end{array}
 \end{equation}
In the above equation, $d{\cal I}$ is the dimensionless invariant phase space for the `2-body' partonic final state  $\bar c + x$ where $x$  treated as   one particle. This formula is also correct for higher order partonic cross sections.

The cross section of the $\bar{D}^*$ is  dependent on both the combination matrix element and the distribution of the light quark as shown in the above equation, which are not calculable by pQCD and we should find ways to extract from more `simple'  experiments. For the  light quark distribution $P(\xi_l)$, it is more easy to be modeled,   inspired by the available data and theory for certain cases. There is another special property is that $P(\xi_l)$ is dependent on the specific `background', i.e., vacuum, comoving partons in a jet, underlying events,  quark gluon plasma or even curved space-time etc.

On the other hand, the combination function  $\tilde{F}(\xi, \xi_l)$ is not intuitive. So the model is difficult to be  constructed.   If the cross section of a  more simple process can be factorized and includes this parameter, it can be extracted from data. The following is an example. Let's see the factorization and the complexity.

It has been pointed out that, in hadronic interaction, the asymmetry of $D$ meson in forward direction can be explained by the the combination of the initial parton with the charm quark produced in the hard interaction. Such a  leading particle effect  has been studied in  \cite{Braaten:2002yt,Braaten:2001bf,Braaten:2001uu,Chang:2003ag}, both in the approximation $m_c \to \infty$. In such an approximation, the light quark has vanishing momentum, hence, qualitatively, the momentum of the $D$ meson is approximately that of the charm quark, so that   not possible to probe the momentum of the light quark, but leaving a non relativistic combination matrix.  On the other hand, in \cite{Collins:1981uw}, the authors also  tried to give the combination matrix elements in the framework of collinear factorization, which is the same framework used in this paper.  The combination matrix elements  there depend on three variables $z_1$, $z_2$, $z_3$, seem not corresponding to the momentum fraction of the valence partons. However, starting  from Equation (4) in \cite{Chang:2003ag}, by taking into account the space-time transition invariance, we get the combination matrix elements with two variables corresponding to the momentum fraction of the charm and the light quarks, which is like those in the above section:
\begin{eqnarray}
\label{uccomme}
& & \int \frac{d^4 k_{c}}{(2 \pi)^4} \frac{d^4 k_{l}}{(2\pi)^4} \delta(\xi-\frac{k^+_{c}}{K^+}) \delta(\xi_l-\frac{k^+_{l}}{K^+}) \int d^4 x_1 d^4 x_2 e^{i k_{c} x_1} e^{-i k_{l} x_2} \nonumber \\
&\times  & <0| \bar q_{k}(0) \frac{\gamma^+}{2 K^+} Q_{l}(x_1))|H_Q>
<H_Q| \bar Q_{i}(0) \frac{\gamma^+}{2 K^+}  q_j(x_2))|0>,
\end{eqnarray}
and
\begin{eqnarray}
\label{uccomme2}
& &\int \frac{d^4 k_{c}}{(2 \pi)^4} \frac{d^4 k_{l}}{(2\pi)^4} \delta(\xi-\frac{k^+_{c}}{K^+}) \delta(\xi_l-\frac{k^+_{l}}{K^+}) \int d^4 x_1 d^4 x_2 e^{i k_{c} x_1} e^{-i k_{l} x_2} \nonumber\\
&\times  & <0| \bar q_{k}(0) \frac{\gamma^5 \gamma^+}{2 K^+} Q_{l}(x_1))|H_Q>
<H_Q| \bar Q_{i}(0) \frac{\gamma^5 \gamma^+}{2 K^+}  q_j(x_2))|0>.
\end{eqnarray}
These two parts of the combination matrix element, i.e., the double-vector part and the  double-pseudo-vector part, should be separated  since the partonic cross sections corresponding to these two parts could be different. For the case that the  quark and the anti-quark  from different sources respectively, these two parts can be put together and only the vector part needs  consideration.

The complexity lies in that,  Equations (\ref{uccomme}, \ref{uccomme2}) are different from that in Equation (\ref{eq:ft}) by  $2 E \delta^3 ({\bf k}_{\bar c}+{\bf k}_l -{\bf K})$,  without the restriction $\xi+\xi_l=1$.  The reason is that in the process in \cite{Chang:2003ag},  the  light quark and the  heavy quark can undergo hard interactions and in principle are not restricted on mass shell.  We can also understand this from a different way. Equations (\ref{uccomme}) and (\ref{uccomme2})  look like the combined distribution of two valence quarks in the heavy meson. In fact, rewriting the $<0|(\cdot \cdot \cdot)_1 |H_Q><H_Q|(\cdot \cdot \cdot)_2|0> $ to the form $<H_Q|(\cdot \cdot \cdot)_2 |0><0|(\cdot \cdot \cdot)_1|H_Q> $, integrating the $\delta$ functions and the exponential functions, taking $|0><0|=1$ in the vacuum saturation approximation, we will get the form of the product of two parton distribution functions, each similar to that defined by  Collins and Soper~\cite{Collins:1981uw}. Then in a parton model at high energy, we can not require the sum of two parton momentum fractions equals one. Hence to get the inputs needed, we should find ways to relate the `restricted' one above with this `unrestricted'.

If the unrestricted matrix elements have been  extracted from experiments, to get  the restricted  combination function, we start from, by denoting the Combination Matrix Element in Equations (\ref{uccomme}, \ref{uccomme2})  as $CME$:
\begin{eqnarray}
\label{CMEeq}
CME&=&\int \frac{d^3 K'}{2E'} 2E' \delta^3({\bf k_c'+k_l'-K'}) \frac{d^3 k_c'}{2E_c'} 2E_c' \delta^3({\bf k_c'-k_c}) \nonumber \\
&\times & \frac{dk_l'}{2E'_l} 2E_l' \delta^3({\bf k_l'-k_l})  CME \nonumber
\\
&=& \int \frac{d^3 K'}{2E'}  \tilde{F}(\xi,\xi_l; \xi+\xi_l).
\end{eqnarray}
In principle, we should solve the integral equation and use the value of $\tilde{F}$ on $\xi+\xi_l=1$ (${\bf K'=K}$) as our inputs.  On the other hand, if the real world is more simple as most models assume,   $\tilde{F}(\xi,\xi_l; \xi+\xi_l)$ peaks around $\xi+\xi_l=1$, i.e.,  two valence quarks on mass shell with ${\bf k_c+k_l \simeq K}$, we can approximately fit  $CME$ as,
\begin{eqnarray}
CME&=&\int \frac{d^3 K'}{2E'} 2E'\prod \limits_{i}\frac{\epsilon_i}{\pi (\epsilon_i^2+({\bf K}-{\bf K'})_i^2)} M^2 f(\xi,\xi_l) \nonumber\\
 &\overset{=}{(\epsilon_i \to 0)} & \int \frac{d^3 K'}{2E'} 2E'\delta^3({\bf K}- {\bf K}') M^2 f(\xi,\xi_l).
\end{eqnarray}
So in the extreme/ideal condition, we can just have $\tilde{F}(\xi,\xi_l)_{\xi+\xi_l=1} \propto \frac{CME}{M^2}$. That is, because  the distribution of $\tilde{F}$ is a narrow peak, we use the average value of it in a reasonably small integral region of $\frac{d^3 K'}{2 E'}$.

Equations (\ref{uccomme}, \ref{uccomme2}) will be applied in other processes and discussed in elsewhere, so that we can have more experiments to extract the combination matrix elements. In principle, when we accumulate enough number of data, especially from more than one process, the integral Equation (\ref{CMEeq}) can be solved. We just mention that we have discussed the combination process preliminarily in $ e^+$  $e^-$ annihilation, where a light quark `fragments' into a heavy meson by combination \cite{Jin:2003vi}. In next section, we will show how to explain data by reasonable physical input.

%==================================================================
\section{$D^*$ in a jet  from proton proton scattering}\label{sec:num}
\begin{figure}[thb]
%\label{cvf}
	\centering
	\begin{tabular}{cccccc}
		\scalebox{0.380}[0.4]{\includegraphics{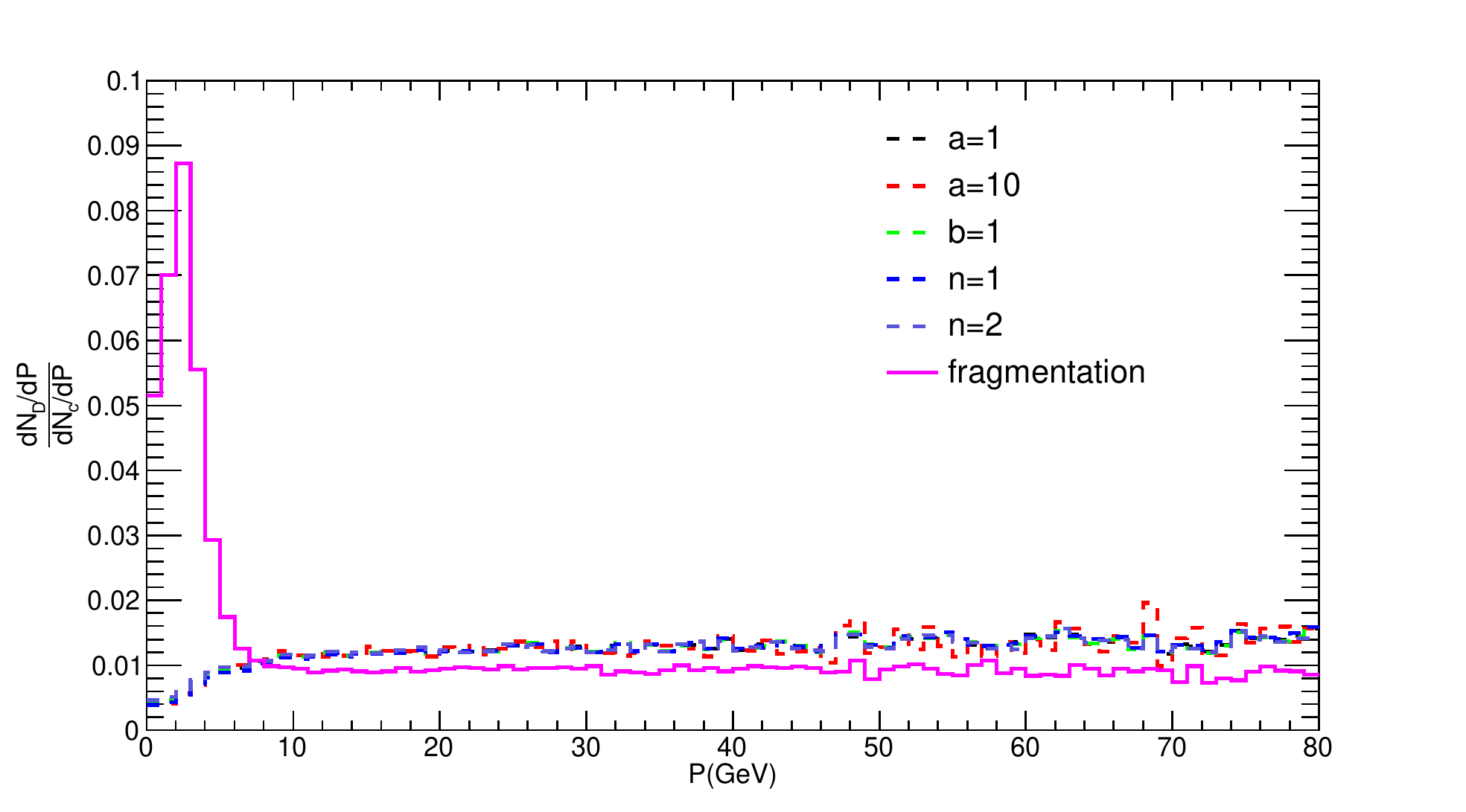}}&
		\scalebox{0.38}[0.4]{\includegraphics{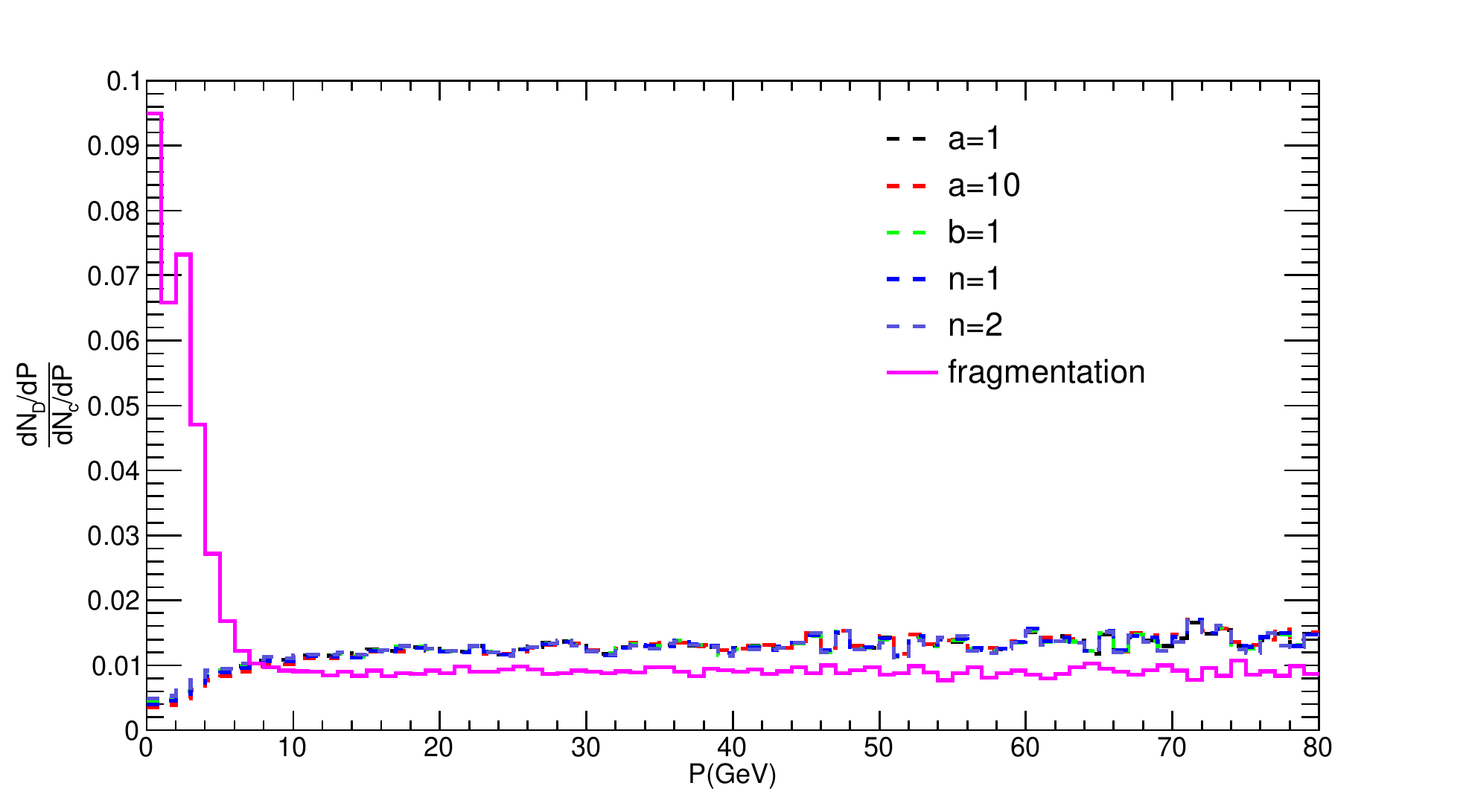}}&\\
		{\scriptsize (a) $\sqrt{s}=7$TeV}&{\scriptsize (b) $\sqrt{s}=14$TeV}
		
	\end{tabular}
	\caption{Comparison of the combination effect to the fragmentation.  We show
the spectrum of the $\bar D^*$ meson rescaled by the spectrum of the charm quark  in pp collision at LHC.
The combination matrix element $\tilde{F}(\xi,\xi_l)=1$, $P(\xi_l)$  as Gaussian  $e^{{-\xi_l}^2/a}$, exponential $e^{-\lambda/b}$ and polynomial  $1/(1+\xi_l)^{n}$. The hot pink curve is for the fragmentation.}
	\label{fig:1}
\end{figure}

To begin with, we first study the phenomenology difference between the conventional fragmentation production and that from the combination production which can be calculated by  Equation~(\ref{cossnum}). The fragmentation can be  calculated by   generators such as  PYTHIA\cite{Sjostrand:2006za}. For consistency, we also use the same matrix element calculation as those in PYTHIA on the partonic 2 to 2 cross sections in  Equation~(\ref{cossnum}). We  take the combination matrix element $\tilde{F}(\xi,\xi_l)$ as a constant, i.e., the free combination without restrictions, to  demonstrate the largest combination effect.  The c.m.s energies are taken to be 7 and 14 TeV. To do the calculations we also have to set the distribution of the parton which to be combined with the charm quark. We tried with several forms of $P(\xi_l)$, e.g., Gaussian  $e^{{-\xi_l}^2/a}$, exponential $e^{-\lambda/b}$ and polynomial  $1/(1+\xi_l)^{n}$. The parameters can be found in Fig.~\ref{fig:1}.

From Fig.~\ref{fig:1}, we see that the difference of the combination from the fragmentation is significant. However, the above result is not a realistic one. The key point is that  when considering the combination effect, the real gluon radiation in the perturbative part have to be taken into account, i.e., the partonic
cross section in Equation~(\ref{cossnum}) must be modified by the summation to  all order real gluon radiation corrections. This is the similar case as considering the fragmentation contributions. In the corresponding physical process, it is the development of the jet. So
  we investigate the prompt $D^{*}$  meson in a charm jet. In this case the perturbative evolution in the jet development should be included, which can be described  by the  parton shower in PYTHIA.

According to the experimental measurement\cite{ATLAS:2011chi}, a function $R(p_{T},z)$ is defined as
  \begin{eqnarray}
\label{jff}
R(p_{T},z)\equiv \frac{N_{D^{*\pm}}(p_{T},z)}{N_{jet}(p_{T})},
\end{eqnarray}
for a certain region set by the  $p_{T}$ and rapidity of the jet, with  $z=|\vec p_{h}\cdot \vec p_{jet}|/\vec p^{2}_{jet}$. For the experimental observed $D^* $, the cross section can be written as
  \begin{equation}
  \label{totxs}
  d \sigma = d\sigma ^F + d \sigma ^C +... ,
  \end{equation}
this means that the total cross section includes the fragmentation contribution $d\sigma ^F$, the combination contribution  $d\sigma ^C$, and other high twist contributions which we assume neglectable in this study. As mentioned above, the fragmentation function is calculated by PYTHIA. For $d\sigma ^C$, since we can not calculate the combination function $\tilde{F}(\xi,\xi_l)$, we have to give some physical input to it. $R(p_{T},z)$  is only measured for  the region of $z>0.3$  by ATLAS~\cite{ATLAS:2011chi}. So we have an extreme try---only considering the charm quark which can not contribute to the data via fragmentation.  Say, we only consider the contribution from the $z_{c}=p_{c}/p_{jet}<0.3$ after the parton shower. The reason is that the fragmentation  always gives a $D^{*\pm}$ meson with momentum smaller than that of the charm quark. That is, $z<z_{c}$ in fragmentation process.

\begin{figure}[tbh]
	\centering
	\begin{tabular}{cccccc}
		\scalebox{0.60}[0.60]{\includegraphics{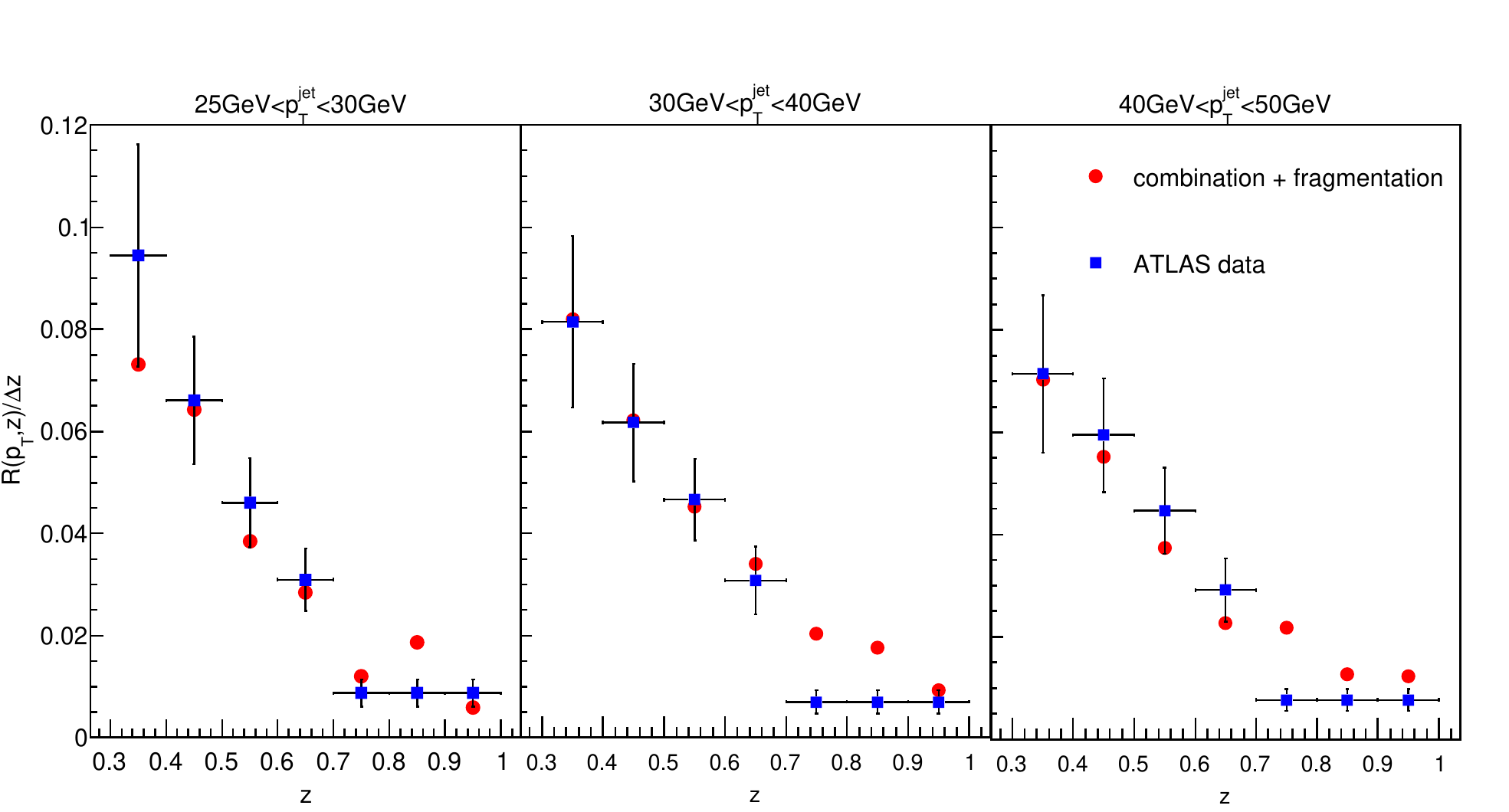}}\\
		\scalebox{0.60}[0.60]{\includegraphics{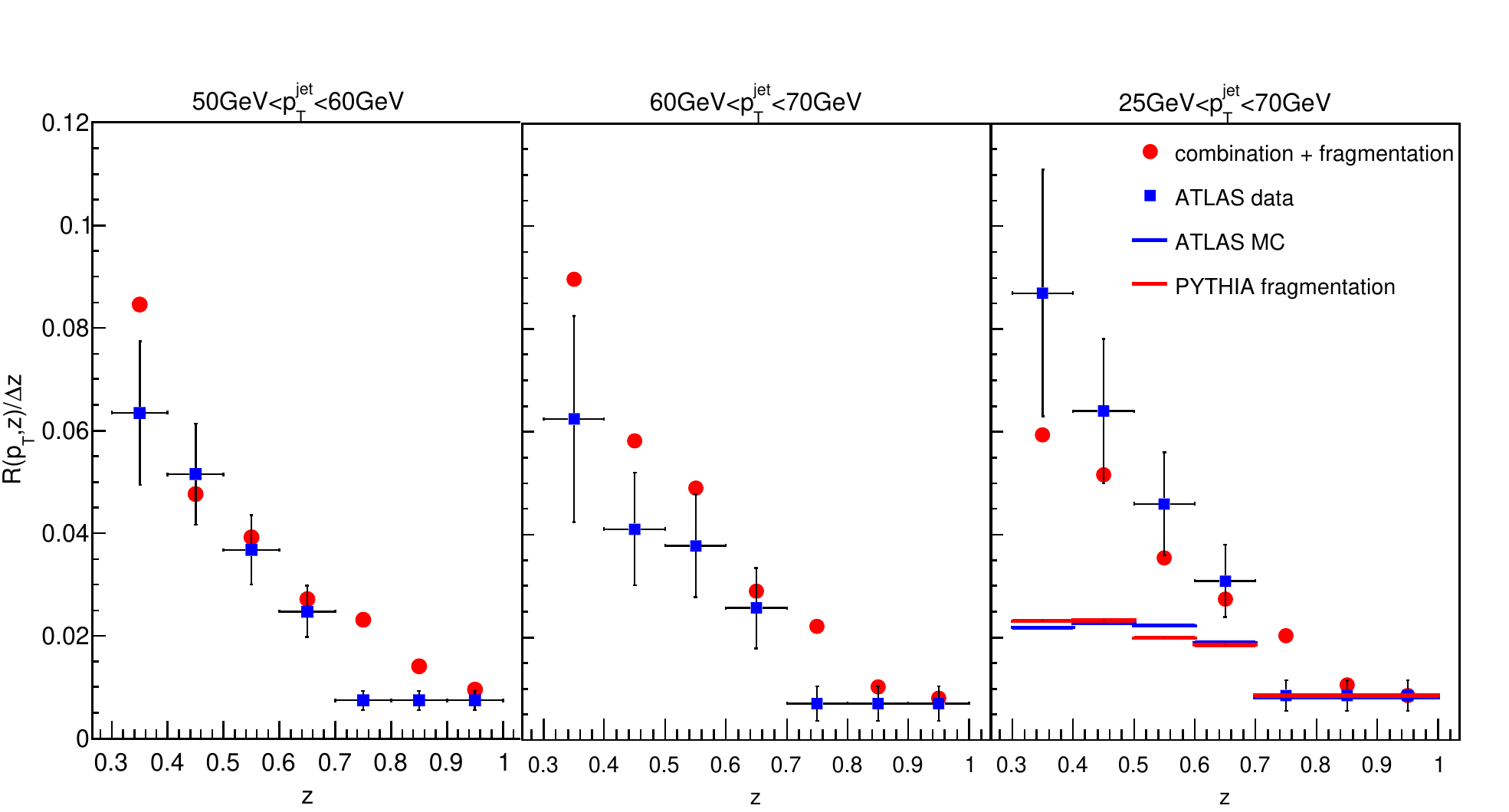}}\\
	\end{tabular}
	\caption{$R(p_{T},z)$ distribution from the summation of both the combination and fragmentation contributions as Equation~(\ref{totxs}) (red disc) compared to the data (blue square) from the ATLAS~\cite{ATLAS:2011chi}.
Here $\tilde{F}(\xi,\xi_l)=4\xi^{2}$ and $P(\xi_l)=e^{-0.1\xi_l^{2}}$.}
	\label{fig:3}
\end{figure}

The distribution of light partons is set to be $P(\xi_l)=e^{-0.1\xi_l^{2}}$ and the  non-zero value of the combinational function is taken to be $\tilde{F}(\xi,\xi_l)=4\xi^{2}$  with $z_{c}<0.3$.  The  $R(p_{T},z)$ distribution of  the $D^{*\pm}$ mesons produced from summation of the  fragmentation  and combination contributions according to Equation~ (\ref{totxs}) are shown in Fig.~\ref{fig:3}. The data \cite{ATLAS:2011chi} are also plotted. In the full $p_{T}^{jet}$ region with $25 <p_{T}^{jet}<70 ~\rm{GeV}$, our results agree with the experimental data. The more detailed comparison in the region of $25<p_{T}^{jet}<30 ~\rm{GeV}$, $30<p_{T}^{jet}<40 ~\rm{GeV}$, $40<p_{T}^{jet}<50 ~\rm{GeV}$, $50<p_{T}^{jet}<60 ~\rm{GeV}$ and $60<p_{T}^{jet}<70 ~\rm{GeV}$,   are also consistent with the data.

\begin{figure}[htb]
	\centering
	\begin{tabular}{cccccc}
		\scalebox{0.60}[0.60]{\includegraphics{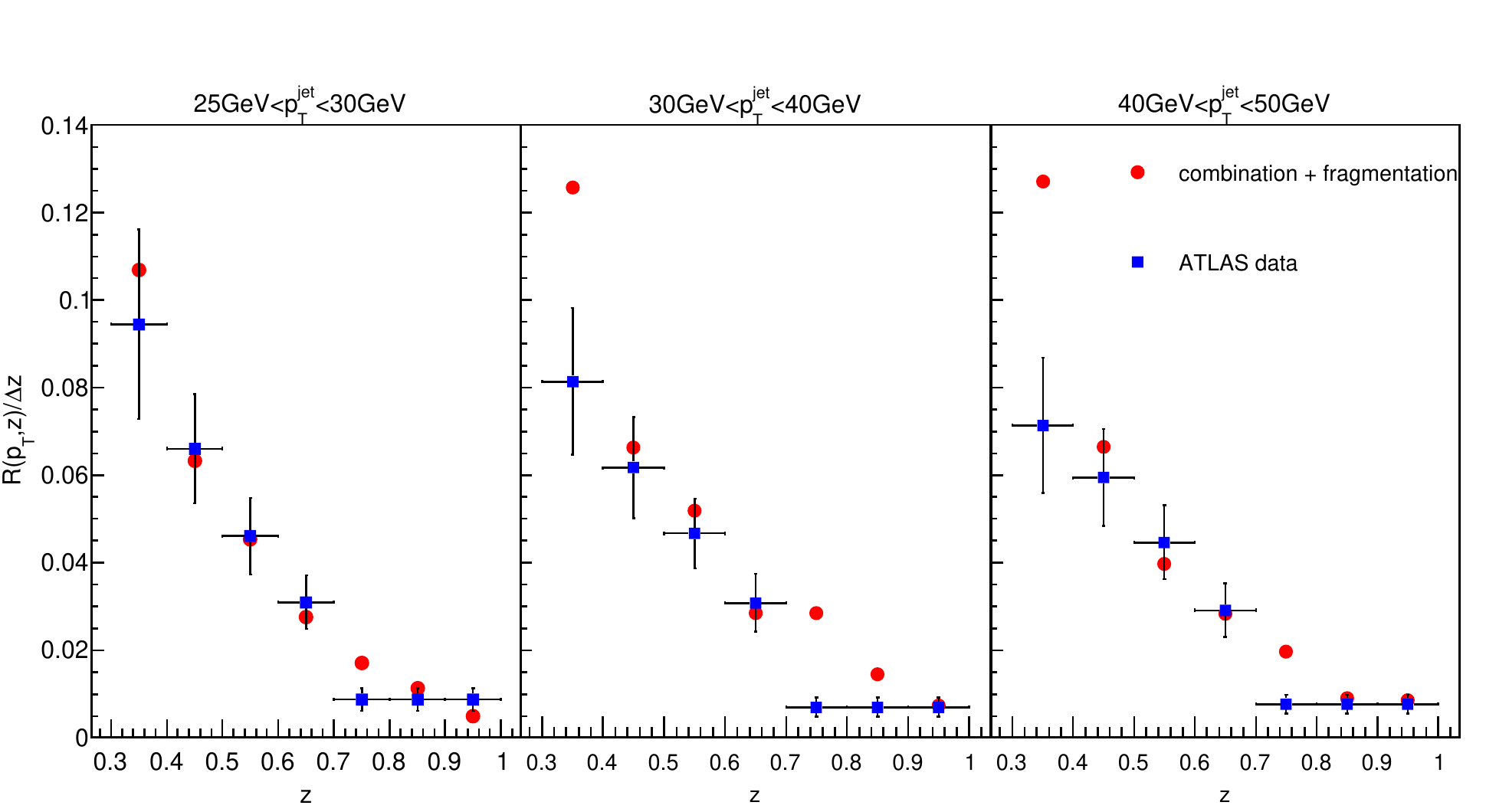}}\\
		\scalebox{0.60}[0.60]{\includegraphics{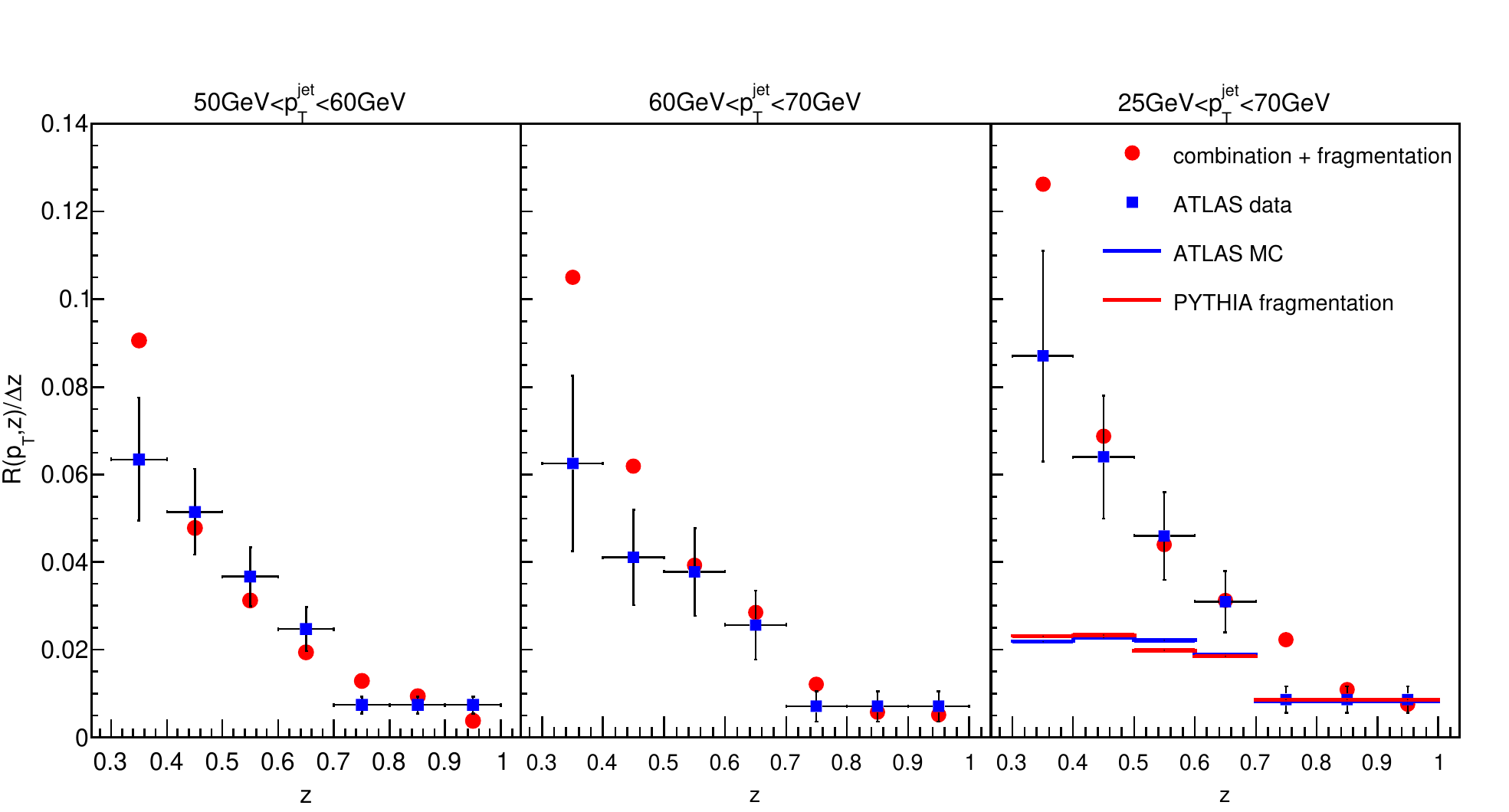}}\\
	\end{tabular}
	\caption{$R(p_{T},z)$ distribution from both the combination and fragmentation contributions (red disc) compared to the data (blue square) from the ATLAS~\cite{ATLAS:2011chi}. Here the combination probability function is taken as $\kappa(z)=\lambda e^{-\lambda z}/(1-e^{-\lambda})$ with $P(\xi_l)=e^{-\xi_l^{2}}$, $\lambda=4.6$. }
	\label{fig:4}
\end{figure}

In the above calculation, we only consider the contribution from  the combination mechanism taking effect  in  the  region of small momentum $0<z_c<0.3$. This is just a na\"ive separation of the contribution from the fragmentation or the combination. However, from the experimental viewpoint, it is $z$  but not the $z_c$ the variable which can be measured from the final hadrons $D^{*\pm}$. To mimic the real process, we propose a   distribution function $\kappa(z)$ to assign the probability of   the combination mechanism. Since the fragmentation mechanism has been tested and applied in various processes, and the theoretical framework is very mature, embedded into the generator,  we can get $\kappa(z)$  via fitting the data. The result is shown in Fig.~\ref{fig:4}. Here we extract the probability distribution function $\kappa(z)=\lambda e^{-\lambda z}/(1-e^{-\lambda})$ with $P(\xi_l)=e^{-\xi_l^{2}}$, $\lambda=4.6$. By this way we can describe the data well.

%********new section
\section{Discussion and application}\label{sec:disc}
Hadron production at the high energy collision provides plenty of phenomena for the study of QCD. In this paper we have derived the factorization formula for the $D^*$ production in the combination process.  As an application, we have summed the combination and fragmentation contributions to compare with the experimental data. With the proper combination function $\tilde{F}(\xi,\xi_l)$ and the light parton distribution function $P(\xi_l)$, the calculations are in agreement with the data.

The above investigations demonstrate the importance of the combination mechanism. The experiments which measure the hadron distribution in a jet show the hadronization process is more complex than a simple fragmentation picture, especially the small $z$ distribution. Such a  phenomenon is also observed in heavy ion scattering as measuring the jet quenching~\cite{ALICE:2015ccw,ALICE:2015vxz,CMS:2017qjw}. These two cases share some similarity such as that  underlying events are important. If we consider the hadronization process as an effective coupling of the quark degree of freedom with the hadronic  one, then the simplest operator relevant is the fragmentation, including two field operators, one is quark field and the other is hadron. The next one to be considered is the  operators including two quark fields and one hadron field. This is the combination contribution. When the combination is taken into account, the  density of the partons will make sense, as shown by our formulations above. From this consideration, one can suspect the larger the partonic density, the more the combination  contributes. So further measurements in higher energies as well as multiplicity-triggered events can give richer phenomenology.

The combination matrix elements are not yet available from data or some non-perturbative calculations, this is why we make trial in the above section. It is obvious that with more and more measurements on the processes  including combination contribution, one can understand more. On the other hand, The quark (re)combination models are of a long history. They can date back to more than three  decades ago. From then on, different kinds of quark (re)combination models  are presented in application to hadron production in various high energy processes, (see, e.g.,~\cite{Jin:2010wg} and references therein). These models work on quark-gluon degree of freedom as well as hadronic one.  So the above derivation and argument can be applied to combination of hadrons or quark clusters as the following example.
 \begin{figure}[t]
	\centering
	\begin{tabular}{cccccc}
		\scalebox{0.50}[0.60]{\includegraphics{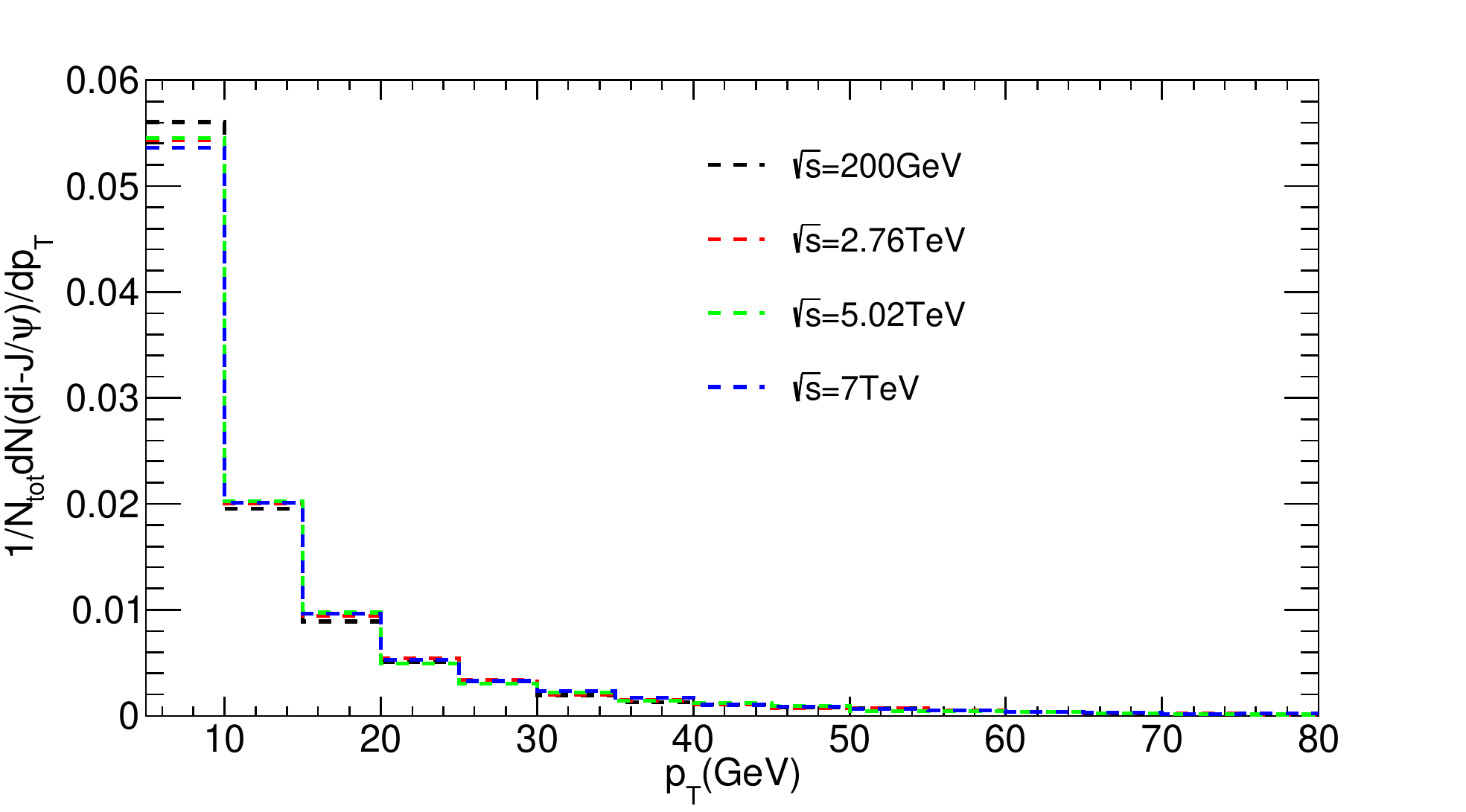}}\\
	\end{tabular}
	\caption{The $J/\psi$ pair transverse momentum spectra of $\sqrt{s}=200$GeV, $\sqrt{s}=2.76$TeV,$\sqrt{s}=5.02$TeV and $\sqrt{s}=7$TeV at pp collision, supposed $P(\xi_l)$ as Gaussian distribution $e^{{-\xi_l}^2}$.}
	\label{fig:5}
\end{figure}

Recently, a new resonance in the invariant mass spectrum of the $J/\psi$ pair~\cite{LHCb:2020bwg} was observed at  the LHCb. The internal structure of such state usually assumes the diquark and antidiquark attracting each other.   We extend the combination formalism to the exotic hadron productions, supposing that one hard $J/\psi$ combines with a soft one  to form a di-$J/\psi$ as heavy quarks do. In other ways, one can also  suppose the combination happen on two pair of diquarks($cc-\bar{c} \bar{c}$). Here we show the $p_{T}$ spectrum of di-$J/\psi$ in Fig.~\ref{fig:5} with $P(\xi_l)=e^{-\xi_l^{2}}$, $\tilde{F}(\xi,\xi_l)=1$ via the combination of two $J/\psi$'s. The distributions are weighted by the total events at the corresponding energy. This indicates an universal study of combination mechanism can be done via various combination processes available.

\section*{Acknowledgement}
This work was supported by the National Natural Science Foundation of China (NSFC) under Grant Nos.~11775130 and~\texttt{11635009}  and Natural Science Foundation of Shandong Province under grant No.~\texttt{ZR2017JL006}.

\bibliographystyle{JHEP}
\bibliography{references1}

\end{document}